# Reflection and Self-Monitoring in Quantum Mechanics

Andrew Mason and Chandralekha Singh

*Department of Physics and Astronomy, University of Pittsburgh, Pittsburgh, PA*

**Abstract.** An assumed attribute of expert physicists is that they learn readily from their own mistakes. Experts are unlikely to make the same mistakes when asked to solve a problem a second time, especially if they have had access to a correct solution. Here, we discuss a case study in which fourteen advanced undergraduate physics students taking an honors-level quantum mechanics course were given the same four problems in both a midterm and final exam. The solutions to the midterm problems were provided to students. The performance on the final exam shows that while some advanced students performed equally well or improved compared to their performance on the midterm exam on the questions administered a second time, a comparable number performed less well on the final exam than on the midterm exam. The wide distribution of students' performance on problems administered a second time suggests that most advanced students do not automatically exploit their mistakes as an opportunity for learning, and for repairing, extending, and organizing their knowledge structure. Interviews with a subset of students revealed attitudes towards problem-solving and gave insight into their approach to learning.

## INTRODUCTION

It is commonly assumed that most students who have made it through an entire undergraduate physics curriculum have learned not only a wide body of physics content but also have picked up the habits of mind and self-monitoring skills needed to build a robust knowledge structure[1]. Instructors take for granted that advanced physics students will learn from their own mistakes in problem solving without explicit prompting, especially if students are given access to clear solutions. It is implicitly assumed that, unlike introductory students, advanced students have become independent learners and they will take the time out to learn from their mistakes.

However, such assumptions about advanced students' superior learning and self-monitoring skills have not been substantiated by research. Very little is known about whether the development of these skills from the introductory level until the time the students finish their highest degree is a continuous process of development or whether there are some discontinuous ``boosts" in this process for many students, e.g., when they become involved in graduate research or when those who become professors ultimately independently start teaching and researching.

Furthermore, investigations in which advanced physics students are asked to perform tasks related to simple introductory physics content do not fully assess their learning and self-monitoring skills[1,2]. Advanced students may have a large amount of "compiled knowledge" about introductory physics and may not need to do much self-monitoring or learning while dealing with introductory problems.

The task of evaluating advanced physics students' learning and self-monitoring skills should involve advanced level physics topics at the periphery of advanced students' own understanding. While tracking the same student's learning and self-monitoring skills longitudinally is an extremely difficult task, taking snapshots of advanced students' learning and self-monitoring skills can be very valuable. Here, we investigate whether students in an advanced quantum mechanics course can avoid making the same mistake twice on their exams without explicit intervention.

At the University of Pittsburgh, honors-level quantum mechanics is a two-semester course sequence which is mandatory only for those students who want to obtain an honors degree in physics. It is often one of the last courses an undergraduate physics major takes. Here, we discuss a study in which we administered four quantum physics problems in the same semester both in the midterm and final exams to students enrolled in the honors-level quantum mechanics. Solutions to all of the midterm questions were available to students on a course website. Moreover, written feedback was provided to students after their midterm performance, indicating on the exams where mistakes were made and how they can be corrected.

## PROCEDURE

The honors-level quantum mechanics course had 14 students enrolled in it, most of whom were physics seniors. The class was primarily taught in a traditional lecture format but the instructor had the students work on a couple of preliminary tutorials that were being developed. Students were assigned weekly homework throughout the fifteen-week semester. In addition, there were two midterm exams and a final exam. The midterm exams covered only limited topics and the final exam was comprehensive. Students had instruction in all relevant concepts before the exams, and homework was assigned each week from the material covered in a particular week. Each week, the instructor held an optional class in which students could ask for help about any relevant material in addition to holding office hours. The first midterm took place approximately eight weeks after the semester started, and the second midterm took place four weeks after the first midterm. For our study, two problems were selected from each of the midterms and were given again verbatim on the final exam along with other problems not asked earlier. The problems given twice are listed in the Appendix.

### Data Collection Procedure

Three of the problems chosen (labeled as problems 1, 2, and 3 for convenience) were those with which several students had difficulty; a fourth problem (labeled as problem 4) which most students found straightforward on one of the two midterm exams was also chosen. The most difficult of the four problems (based upon students' performance) was problem 3, which was also assigned as a homework problem before the midterm exam. It was perceived by students to be more abstract in nature than the other problems. The easiest of the four problems, problem 4, was an example that was solved within the assigned textbook. The students had access to the homework solutions and midterm problems. Thus, students had the opportunity to learn from their mistakes before they encountered the four problems selected from the midterm exams on their final exam (as noted earlier two problems were selected from each midterm).

A scoring rubric was developed jointly with Yerushalmi and Cohen[3,4,5] to assess how well the students in introductory physics courses diagnose their mistakes when explicitly prompted to do so. This rubric was adapted to score students' performance on each of the four quantum mechanics problems on both the midterm and final exams. The scoring was checked independently by another scorer and at least 80% agreement was found on the scoring for each student on each problem in each attempt (on midterm and final exams). Students were rewarded for correctly identifying and employing physical principles as well as for their presentation and problem-solving skills. Here we summarize the findings based on the performance on the physics scores obtained using the rubric.

To get a better insight, in-depth interviews lasting 1-1.5 hours were conducted with four paid student volunteers from the group of 14 students in the following semester within the first two months using a think-aloud protocol[6]. The goal of these interviews was to learn about students' attitudes and approaches towards problem solving and learning and to better understand their thought processes as they attempted to solve the 4 problems chosen for the study again during the interview. Three of the four interviewed students were enrolled at that time in the second semester course in honors-level quantum mechanics. The fourth student had graduated in the fall semester and was performing research with a faculty member. During these interviews, we first asked students about their approaches and strategies for problem solving and learning and asked them to solve the same four problems again while thinking aloud. We did not disturb them initially when they answered the questions and only asked for clarification of points after the student had answered the questions to the best of his/her ability. These delayed interviews also provided an opportunity to understand how well students had retained relevant knowledge after the semester was over and could retrieve it a couple of months later to solve the problems. Two shorter interviews were conducted later with two additional students which mainly focused on students' attitudes and approaches to learning due to the time constraints.

**TABLE 1.** Average midterm and final exam scores for each student including all four problems. An asterisk indicates a student who was interviewed.

| Student | 1 | 2 | 3* | 4 | 5 | 6* | 7 | 8* | 9* | 10* | 11* | 12 | 13 | 14 |
|---|---|---|---|---|---|---|---|---|---|---|---|---|---|---|
| $m_i$ | 62 | 60 | 93 | 58 | 41 | 88 | 40 | 83 | 93 | 56 | 80 | 100 | 54 | 19 |
| $f_i$ | 56 | 17 | 72 | 34 | 39 | 47 | 63 | 97 | 93 | 47 | 97 | 100 | 64 | 11 |

# RESULTS

Overall, the midterm average score of all students was 66% on all four problems and 57% on the three difficult problems (omitting the "easy" problem). The average final exam score of all students was 60% on all four problems and 53% on the three difficult problems. Thus, the students' average final exam performance on these problems is slightly worse than their performance on the midterm exams. Before we focus at the change in each student's performance from the midterm exam to the final exam on problems given a second time, we note that this lowering of the average score in the final exam compared to the midterm exams suggests that the assumption that the senior-level physics majors will automatically learn from their mistakes may not be valid.

Table 1 contains each student's average score on the four problems for both midterm ($m_i$) and final exam ($f_i$) attempts. It is clear from Table 1 that some students did well both times or improved in performance but others did poorly both times or deteriorated on the final exam. Students struggled the most on problem 3 both in the midterm and final exams, and regressed the most from midterm to final exam on problem 2.

In interviews, students frequently noted that they did not expect a problem they had encountered on a previous exam to occur on the final exam. This expectation often resulted in a more careful review of homework assignments than previous exams. The following comments from students, which pertain to general study habits, exhibit this trend:

*"If I make mistakes in the homework, I look at the TA's solutions carefully because I know those problems can show up in the exams. But if I make a mistake in the midterm exam, I won't be so concerned about what I did wrong because I don't expect those questions to show up in the final exam. Also, if I don't do well on the exam, I don't feel like finding out what I did wrong because reading my mistake again would just hurt me again, and I don't want anything to ruin the after-exam happy time."* (student 8)

*"When I make mistakes I always look back at the work to see where I erred. In most cases I will be more careful in looking over homework than past exams as far as studying purposes go."* (student 9)

In terms of performance on specific problems, the four students who were asked to solve the problems again in the interview had mixed performance on the four problems. They all struggled to some extent on problem 3, ranging from forgetting details about a Taylor expansion used in the problem's solution to being unable to even start the problem. Several complained that they did not anticipate this problem on the exams and therefore did not study it as

**FIGURE 1.** Student 3's midterm and final exam answers for problem 1. The midterm solution is correct but the final exam solution is incorrect and shows irrelevant work.

thoroughly. For example, student 3 claimed that the required proof didn't seem very physical:

*"...you know, sometimes, really mathematical problems... I mean, this isn't terribly mathematical but sometimes problems like this just seem like 'oh, this is just a math thing.' You know [the book] or even the professor, maybe prior to the problem doesn't tell you the importance of the problem, like what it really demonstrates. I remember [the professor] did [afterwards]....But [the book] doesn't say anything, it just says, 'Do it.' I remember thinking, at the time I thought it was just something to make me stay up another hour and a half. And then, you know, it was on the test and I thought, 'I should've paid more attention to it,' and then it was on the final and I thought, 'jeez, I really should've paid attention to it.'"*

Student 10 also noted that he did not perceive problem 3 to be important:

*"It was just one of the problems in the homework. It was never mentioned previously or after, so I didn't assign much importance to it in my head as far as studying goes to it."*

Three students (with student 11 being the exception) struggled with problem 1. The difficulties were specifically around the fact that Dirac notation had not been mastered, and the regression from the midterm attempt to the final exam attempt by two of the students suggests they may have memorized procedures related to Dirac notation in order to "get by". Student 3, whose work on the midterm and final exam for problem 1 is given in Figure 1 (he solved the problem correctly on midterm but did not know how to solve it in the final exam), noted:

*"So the real problem that I have is, um… I never quite mastered the whole bra and ket notation. This is what really is hanging me up here, probably if I had, I wouldn't… be hung up, because I remember it being like 4 lines to do this. I remember it not being a complicated thing." (student 3)*

Student 10 also stated the transition from integral notation to Dirac notation was not clear to him. In contrast, student 11 had no trouble with problem 1:

*"Well, we learned that in particular. It was proven to us in like, three different ways. I remember the page in Griffiths now…"*

For problem 2, students 3, 6, and 10 had a difficult time figuring out what steps were needed to solve the problem. Student 6 displayed common errors such as confusing energy eigenstates with position eigenstates, and said that he failed to learn from mistakes on his midterm exam attempt because his solution, while incorrect, gave an answer that seemed correct (i.e. arriving at an answer slightly different from the one that had to be proved). On the other hand, student 10 simply could not remember enough of the problem solution in order to replicate it:

*"…just from my memory, like, there's just too many holes and stuff because I haven't looked at it or thought about it in a while…"*

We note that with the exception of forgetting some details about the Taylor expansion on problem 3, student 11 displayed excellent physical and mathematical skills. He cited his habit of reflecting on his mistakes as well as his double major in mathematics and physics as reasons for his success.

## DISCUSSION

We find that students' average performance in the final exam on the problems that were given a second time was not significantly better than the average performance on those problems on the midterm exams. While some students improved, others deteriorated. We suggest that many advanced physics students do not routinely exploit their mistakes in problem solving as a learning opportunity. Our study suggests that many advanced physics students may be employing inferior learning strategies, e.g., "cramming" before an exam and selective memorization of content based upon their expectation that those problems are likely to show up on the exam; most do not give a high priority to building a robust knowledge structure. Prior research shows that introductory physics students benefit from explicit interventions to help them develop useful learning and self-monitoring skills[3,4,5,7,8]. Similar explicit interventions involving formative assessment might also prove useful in advanced courses and can help advanced physics students in developing habits of mind[9].

## ACKNOWLEDGMENTS

We thank F. Reif, R. Glaser, E. Yerushalmi, and E. Cohen for useful discussions. We thank NSF for awards PHY-0653129 and 055434.

## APPENDIX

Here are the exam questions used in this study. Supplementary materials (e.g. formula sheets) are not included for lack of space.

Problem 1) The eigenvalue equation for an operator $\hat{Q}$ is given by $\hat{Q}|\psi_i\rangle = \lambda_i |\psi_i\rangle$, $i = 1,…, N$. Find an expression for $\langle\psi|\hat{Q}|\psi\rangle$, where $|\psi\rangle$ is a general state, in terms of $\langle\psi_i|\psi\rangle$.

Problem 2) For an electron in a one-dimensional infinite square well with well boundaries at x=0 and x=a, measurement of position yields the value $x = a/2$. Write down the wave function immediately after the position measurement and without normalizing it show that if energy is measured immediately after the position measurement, it is equally probable to find the electron in any odd-energy stationary state.

Problem 3) Write an expression to show that the momentum operator $\hat{P}$ is the generator of translation in space. Then prove the relation. (Simply writing the expression is not sufficient…you need to prove it.)

Problem 4) Find the expectation value of potential energy in the $n^{th}$ energy eigenstate of a one dimensional Harmonic Oscillator using the ladder operator method.